\documentclass[letter,traditabstract,longauth]{aa}
\usepackage{graphicx}
\usepackage{color}
\usepackage{txfonts}

\usepackage{natbib} 

\begin{document}
   \title{A high resolution line survey of IRC\,+10216
    with \textit{Herschel}/HIFI\thanks{Herschel is an ESA space observatory with 
   science instruments provided by European-led Principal Investigator consortia and with 
   important participation from NASA}}
   \subtitle{First results: Detection of warm silicon dicarbide (SiC$_2$)}
   \authorrunning{J. Cernicharo et al.}
   \titlerunning{SiC$_2$ in IRC\,+10216}

\author{
Jos\'e Cernicharo\inst{1},
L.B.F.M. Waters\inst{2,3},
Leen Decin\inst{2,3},
Pierre Encrenaz\inst{4},
A.G.G.M. Tielens\inst{5},	
Marcelino Ag\'undez\inst{1,6}, 
Elvire De Beck\inst{3},
Holger S.P. M\"uller\inst{7},
Javier R. Goicoechea\inst{1},
Michael J. Barlow\inst{8},
Arnold Benz\inst{9},
Nicolas Crimier\inst{1},
Fabien Daniel\inst{1,4},
Anna Maria Di Giorgio\inst{10},
Michel Fich\inst{11},
Todd Gaier\inst{12},
Pedro Garc\'{\i}a-Lario\inst{13},
Alex de Koter\inst{2,14},
Theo Khouri\inst{2},
Ren\'e Liseau\inst{15},
Robin Lombaert\inst{3},
Neal Erickson\inst{16},
Juan R. Pardo\inst{1},
John C. Pearson\inst{12},
Russell Shipman\inst{17},
Carmen S\'anchez Contreras\inst{1},
David Teyssier\inst{13}
 }

\institute{
Departamento de Astrof\'{\i}sica, Centro de Astrobiolog\'{\i}a, CSIC-INTA,
Ctra. de Torrej\'on a Ajalvir km 4, Torrej\'on de Ardoz, 28850 Madrid, Spain\\
\email{jcernicharo@cab.inta-csic.es}
\and
Astronomical Institute Anton Pannekoek, University of Amsterdam, Science Park XH, Amsterdam, The Netherlands 
\and
Instituut voor Sterrenkunde, Katholieke Universiteit Leuven, Celestijnenlaan 200D, 3001 Leuven, Belgium
\and
LERMA and UMR 8112 du CNRS, Observatoire de Paris, 61 Av. de l'Observatoire, 75014 Paris, France
\and
Leiden observatory, Leiden University, PO Box 9513, NL-2300 RA Leiden, The Netherlands
\and
LUTH, Observatoire de Paris-Meudon, 5 Place Jules Janssen, 92190 Meudon, France
\and
I. Physikalisches Institut, University of Cologne, Germany
\and
Dept of Physics \& Astronomy, University College London, Gower St, London WC1E 6BT, UK
\and
Institute of Astronomy, ETH Zurich, 8093 Zurich, Switzerland
\and
INAF - Istituto di Fisica dello Spazio Interplanetario, Area di Ricerca di Tor Vergata, via Fosso del Cavaliere 100, I-00133 Roma, Italy
\and
Department of Physics and Astronomy, University of Waterloo, Waterloo, ON Canada N2L 3G1
\and
Jet Propulsion Laboratory, 4800 Oak Grove Drive, MC 168-314, Pasadena, CA 91109  U.S.A.
\and
European Space Astronomy Centre, ESA, P.O. Box 78, E-28691 Villanueva de la Ca\~nada, Madrid, Spain
\and
Astronomical Institute, Utrecht University, Princetonplein 5, 3584 CC, Utrecht, The Netherlands 
\and
Department of Radio and Space Science, Chalmers University of Technology, Onsala Space Observatory, 439 92 Onsala, Sweden
\and
University of Massachusetts, Astronomy Dept., 710 N. Pleasant St., LGRT-619E, Amherst, MA 01003-9305  U.S.A.
\and
SRON Netherlands Institute for Space Research, Landleven 12, 9747 AD Groningen, The Netherlands
}
\date{Received / accepted}
\abstract{ 

We present the first results of a high-spectral-resolution 
survey of the carbon-rich evolved star IRC\,+10216 that was carried out with the 
HIFI spectrometer onboard Herschel. This survey covers all HIFI bands, with 
a spectral range from 488 to 1901\,GHz. In this letter we focus on the band-1b spectrum, 
in a spectral range $554.5-636.5$\,GHz, where we identified 130 spectral features with 
intensities above 0.03\,K and a signal--to--noise ratio $>$5. Detected lines arise from 
HCN, SiO, SiS, CS, CO, metal-bearing species and, surprisingly, silicon dicarbide  (SiC$_2$). 
We identified 55 SiC$_2$ transitions involving energy levels 
between 300 and 900\,K. By analysing these rotational lines, we conclude that SiC$_2$ is 
produced in the \textit{inner} dust formation zone, with an abundance of $\sim$2$\times$10$^{-7}$ 
relative to molecular hydrogen. These SiC$_2$ lines have been observed for the first time in 
space and have been used to derive an SiC$_2$ rotational temperature of $\sim$204\,K and a 
source-averaged column density of $\sim$6.4$\times$10$^{15}$\,cm$^{-2}$. Furthermore, the high quality of the HIFI data 
set was used to improve the spectroscopic rotational constants of SiC$_2$.
}
\maketitle
%

\section{Introduction}
Circumstellar envelopes (CSEs) of evolved stars foster a remarkably complex chemistry.
The reactions behind molecular synthesis, how they vary 
from the inner to the outer 
layers of the expanding envelope, their relationship to dust grain formation, and 
the chemical evolution from the 
asymptotic giant branch (AGB) to the post AGB-phases, are all important 
questions that still need answers \citep{Cernicharo2004, Herpin2001,Herpin2002}. 
Observations at different wavelength ranges are clearly required with
this purpose in mind.
The atmosphere and 
the inner dust condensation regions are best probed by infrared (IR) ro-vibrational lines 
or by high-$J$ lines in the ground and vibrationally excited states of abundant species such as CO, HCN, and 
SiS \citep{Fonfria2008,Cernicharo1996a,Decin2010,Patel2009}. The outer and colder layers of the CSE 
are best probed by molecular rotational lines at millimeter wavelengths \citep{Cernicharo2000,Kemper2003,Agundez2006,
He2008}. 
Based on a large set of high spectral resolution data, the physical conditions of the 
different regions of the CSE can be constrained  by providing a complete picture of the object; 
see, e.g., \citet{Agundez2009,Cernicharo2000,Pardo2007,Patel2009}. 
Of particular interest are the high angular
resolution observations by Lucas et al. (1995, 1997) and \citet{Patel2009} of
the innermost region of the envelope.                                                                                                    

IRC\,+10216 (CW Leo) is one of the brightest infrared sources in the sky, 
making it an ideal target to observe with the 
$\textit{Herschel Space Observatory}$ \citep{Pilbratt2010}. 
Most of the molecules detected in this source are heavy carbon-chain radicals \citep{Cernicharo1996b}, 
metal-bearing species \citep{Cernicharo1987}, anions (see, e.g., \citet{McCarthy2006}; \citet{Cernicharo2008} 
and references therein), and many diatomic and triatomic molecules.
The HIFI instrument \citep{deGraauw2010} provides both 
a high spectral resolution and a wide spectral coverage. The first is necessary for resolving 
the complex kinematics characteristic of IRC\,+10216, 
allowing us to distinguish between the contribution from the inner acceleration zone 
\citep{Fonfria2008,Cernicharo2010a,Decin2010} 
and from the expanding envelope where the 
gas reaches its terminal velocity
\citep{Agundez2009}. 
The wide spectral 
coverage is mandatory for a complete inventory of lines 
to study the chemical 
content and molecular excitation in detail.
In this \textit{Letter} we present the preliminary results of a full line survey of 
IRC\,+10216 taken with HIFI 
between 480 and 1900\,GHz, and highlight the range 554-636\,GHz. We focus 
on silicon dicarbide, SiC$_2$ \citep{Thaddeus1984}, a triatomic molecule that, together 
with HCN, is the main contributor to the forest of lines observed in the submillimeter 
and far-infrared domains.

\begin{figure*}
\includegraphics[angle=-90,width=18.0cm]{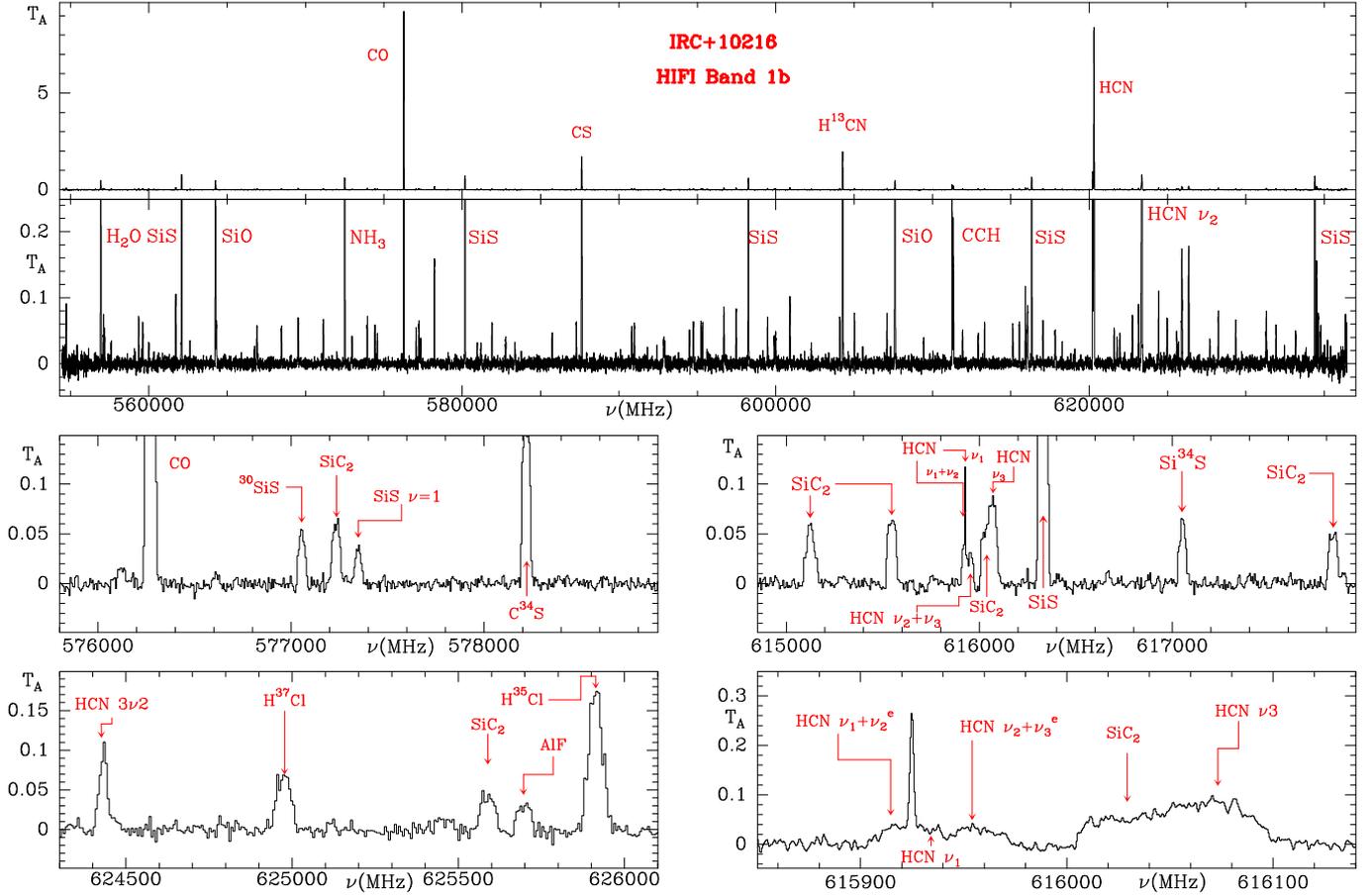}
\caption{Spectra of IRC\,+10216 observed with HIFI band 1b. The two 
upper panels present the complete 
spectrum on two different intensity scales. The panels below show different 
3\,GHz wide ranges of the survey. All data have been smoothed to a spectral resolution of 
2.8\,km\,s$^{-1}$ except for the right bottom panel, which shows the spectrum around several vibrational 
lines of HCN with the nominal WBS resolution (1.1 MHz, $\simeq$0.5 km\,s$^{-1}$). 
} \label{fig-survey}
\vspace{-0.1cm}
\end{figure*}

\section{Observations and data reduction}
HIFI observations \citep{deGraauw2010} were carried out May 11-15 2010. 
In-orbit instrument performances are described in 
detail by \citet{Roelfsema2010}. This survey uses all HIFI bands, providing a frequency coverage between 488 and 1901\,GHz. 
We present the spectral scan of IRC\,+10216 in band 1b (554.5-636.5\,GHz), corresponding to the data of OBSID1342196414.  
The main-beam antenna efficiency at 600\,GHz is 0.75 and the half-power beam width is 36". The data were
taken in double beam-switching mode at a frequency resolution of 1.1\,MHz. The
total integration time per frequency setting was 30.6 seconds (ON+OFF). 
The measured rms noise for each individual frequency 
setting is between 25-40\,mK depending on the frequency.
Averaging all scans (see below) brings 
the measured rms noise down to values between 7 and 12\,mK. 
The data were processed using the standard \textit{Herschel} pipeline up to Level 2, 
providing fully calibrated spectra of the source. We then analyzed them with the GILDAS-CLASS90 
software\footnote{See http://www.iram.fr/IRAMFR/GILDAS for more information about GILDAS softwares.}.

The single side band spectrum of receiver 1b was obtained through a standard manual procedure. We first removed 
spurious features (one) and then compared 
all scans one by one for a given frequency to remove unwanted lines from the corresponding image 
side band by blanking out the corresponding channels. 
In the few cases where image band lines were blended with signal band lines, a fit was performed to the blended lines to separate the contribution from each band. 
The image band emission was then subtracted from the signal band spectrum. This procedure was repeated twice in order to ensure a complete 
cleaning of lines from the image side band at each frequency setting. 
Once the deconvolution was done for one of the receivers, the other receiver was treated automatically 
by blanking out the same frequency blocks in each individual scan. Linear baselines were subtracted and all 
1456 single side band individual scans (728 per receiver) merged. 
The calibration and the double side band gain ratio was checked against the strongest lines, and found to be 
consistent 
within 2-4\%.
The final spectrum with a spectral resolution 
of 2.8\, km\,s$^{-1}$ is shown in Fig.\,1.

\section{Results}
While ground-based observations have provided much insight into the chemical inventory, 
the unprecedented spectral 
coverage of HIFI will yield many lines arising from a wide variety of species that together will greatly expand 
our understanding of the density, 
temperature, and dynamical structure of the ejecta and the processes driving the molecular complexity of 
these chemical ``smokestacks'' 
of the galaxy, from the dust forming zones to the radical and photodominated outer regions.
In this work we present the first 
high spectral resolution line survey towards an evolved star in the submillimeter and far--IR domains. 
Figure\,1 shows around 130 well-detected lines with intensities exceeding 30\,mK. All of them, except one, 
can be easily assigned to lines of CO 
(including $\nu$=1; see Patel et al. 2009b), HCN and H$^{13}$CN (in several vibrational states), SiO, SiS, and CS
(including its isotopologs and vibrationally excited levels), AlF, and AlCl. 
All these species have been previously detected by several authors \citep{Cernicharo2000,
Cernicharo1987,Turner1987,Lucas1989,Lucas1995,Lucas1997,Fonfria2006,Patel2009}.
Of particular interest are the 
$J$=1-0 lines of HCl and H$^{37}$Cl, shown in the bottom left panel of Fig.\,1. \cite{Cernicharo2010b} 
have recently detected these species towards IRC\,+10216 using the PACS and SPIRE instruments 
\citep{Poglitsch2010,Griffin2010}, and the 
present results definitively confirm this detection. 
From the HIFI data we derive X($^{35}$Cl)/X($^{37}$Cl)=2.5$\pm$0.2, consistent with the value derived from the metal-bearing species 
NaCl, AlCl, and KCl isotopologs \citep{Cernicharo1987,Cernicharo2000}, and very close to
the Solar System value of 3 and the one
derived with HIFI in the ISM \citep{Cernicharo2010c}.

\begin{figure}
\vspace{-.8cm}
\hspace{-.05\linewidth}\includegraphics[angle=0,width=1.2\linewidth]{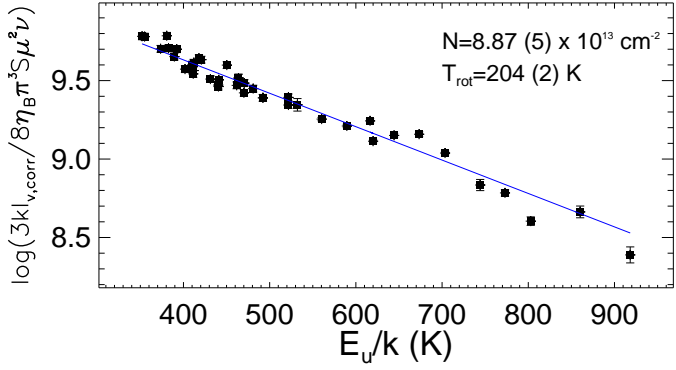}
\caption{Rotational diagram for the observed SiC$_2$ lines. The rotational temperature 
is 204\,K and the beam averaged column density of SiC$_2$ is 8.9$\times$10$^{13}$\,cm$^{-2}$.} \label{fig-rotation}
\vspace{-0.1cm}
\end{figure}

HCN is known to show maser emission in its bending mode and in high vibrationally excited 
states \citep{Guilloteau1987,Lucas1989,Schilke2000,Schilke2003}. Except for one of the 
HCN $J$=6-5 lines, all HCN lines in band 1b seem to be thermal in nature. Figure\,1 (right bottom panel) shows 
that, in addition to a pedestal of thermal emission due to $\nu_1+\nu_2^{1e}$, $\nu_2+\nu_3^{1e}$ and $\nu_1$ 
vibrational levels, there is a clearly distinguished narrow feature that is likely a 
maser line. It can be assigned to $\nu_1$ or to the $\nu_1+\nu_2^{1e}$ 
vibrational level \citep{HCN-high-v_rot_2003}. However, 
without a full detailed treatment of the HCN radiative transfer, it is difficult to assign it to one or the other. 
The $\nu_1+\nu_2^{1f}$ $J$=6-5 line does not show such any equivalent feature. From a very quick look at 
the line survey 
data in other HIFI bands, it appears that the number of HCN masers is quite impressive
and does not have any systematic pattern 
within a given vibrational level. It is worth noting that the $\nu_1$ and $\nu_3$ $J$=1-0 lines of HCN have been observed 
with the IRAM 30m telescope at the level of a few mK. 
The HIFI data show the importance of increasingly higher frequencies (hence 
higher Einstein coefficients and line opacities) to observe the warm layers of CSEs. In particular, 
\citet{Cernicharo2010a} have found many lines of HCN $J$=3-2 from vibrational levels up to 10000\,K. The $J$=6-5 lines 
of these vibrationally excited levels (up to 6000-7000\,K) also appear in the HIFI line survey of 
band 1b and will be presented and analyzed in a forthcoming paper. Other species such as H$_2$O 
and NH$_3$ are analyzed within the HIFISTARS Guaranteed Time Key Program
\citep{Bujarrabal2010}.
In addition to the lines and species presented 
above, we clearly detect 41 spectral features that belong 
to 55 rotational transitions of SiC$_2$, of which 14 are doublets at the same 
frequency. SiC$_2$ emission reaches its maximum at millimeter wavelengths 
in a shell with a radius of 15$^{\prime\prime}$ like the radicals C$_n$H \citep{Guelin1993,Lucas1995}. 
However, interferometric and single-dish observations of 
this molecule also indicate a large abundance in the inner envelope 
(Gensheimer et al. 1995; Lucas et al. 1995; Cernicharo et al. 2000;
Agundez 2009; Patelet al 2009). 
Table\,\ref{SiC2-lines}
(see appendix\,A in the electronic version of the paper) provides the observed parameters 
for the detected SiC$_2$ lines. All of them arise from levels 
involving energies between 300-900\,K and have flat-topped profiles 
(see Fig.\,1 \&4), therefore indicating that they are formed in the inner, warm layers of the 
CSE. We found that the predicted frequencies for these lines were in error by several MHz 
so we provide a new set of rotational constants  in Appendix\,A.

\section{Discussion}
The observed SiC$_2$ line intensities (see Table\,\ref{SiC2-lines}) were used to determine the 
rotational temperature T$_{\rm rot}$ and the column density N(SiC$_2$) of the emitting gas. The rotational 
diagram in Fig.\,2 yields T$_{\rm rot}$=204$\pm$2\,K and N(SiC$_2$)=8.9(1)$\times$10$^{13}$\,cm$^{-2}$ (averaged 
over the beam of HIFI at 636\,GHz). The relatively high rotational temperature implies 
that the excitation region of the considered emission lines is smaller than about 2$^{\prime\prime}$ in 
radius. This region corresponds to a kinetic temperature T$_K$$>$100\,K, according to the expected T$_K$ radial profile (see,
e.g., \citealt{Fonfria2008} and Fig.\,3). Consequently, the source-averaged column density of SiC$_2$ is 
6.4$\times$10$^{15}$\,cm$^{-2}$. Because of the high beam dilution, the regions inward of 0.5$^{\prime\prime}$ 
do not contribute to the observed SiC$_2$ emission. Assuming a mass-loss rate of 2$\times$10$^{-5}$\,M$_\odot$\,yr$^{-1}$ 
and an expansion velocity of 14.5\,km\,s$^{-1}$, the total column density of H$_2$ in the 0.5$-$2$^{\prime\prime}$ 
region is 3$\times$10$^{22}$\,cm$^{-2}$, and the abundance of SiC$_2$ relative to H$_2$ is 2.9$\times$10$^{-7}$.

The inner envelope abundance derived for SiC$_2$ 
(2.9$\times$10$^{-7}$ from the rotational diagram and 2$\times$10$^{-7}$ from 
the radiative transfer model) is in good agreement with the value computed under thermochemical 
equilibrium ($\sim$ 2$\times$10$^{-7}$ from the photosphere up to $\sim$ 4 R$_*$, using the code described 
in \citealt{Tejero1991}), and somewhat higher than the values calculated from non-equilibrium 
chemical modeling of the inner envelope (10$^{-9}$$-$10$^{-8}$; \citealt{Willacy1998}). Previous studies
based on millimeter-wave interferometric observations have derived values for the inner component of SiC$_2$ 
of 5$\times$10$^{-7}$ \citep{Lucas1995} and $<$ 5$\times$10$^{-8}$ \citep{Gensheimer1995}, in reasonable agreement with 
our derived values. Furthermore, these observational studies also find an enhancement of the SiC$_2$ 
abundance, up to $\sim$ 10$^{-6}$ relative to H$_2$, in the outer envelope of IRC\,+10216, implying that additional formation 
routes must be at work in these outer regions. \citet{MacKay1999} studied the chemistry of silicon in IRC\,+10216 and were unable 
to find any reaction that could enhance the abundance of this species in the outer envelope.
Our chemical model predicts such an abundance enhancement (see below).


\begin{figure}
\includegraphics[angle=0,width=8.5cm]{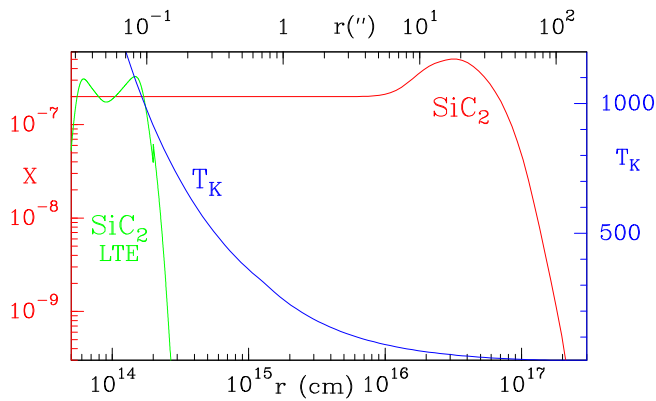}
\caption{Abundance of SiC$_2$, X, derived from the chemical model described in the text (red line) and in 
thermodynamical equilibrium (green line).  The blue line shows the kinetic temperature, T$_K$, of the gas. 
The axis shows the distance to the star in cm (bottom) and the angular distance (top) as seen from the Earth (d=120pc).} \label{fig-abundance} \vspace{-0.1cm}
\end{figure}

To get a more reliable estimate of the abundance of SiC$_2$ in the inner regions and for insight into the peculiar 
chemistry of this species from the inner to the outer layers of the CSE, we carried
out radiative transfer and chemical modeling calculations. We assumed a mass loss rate of 2$\times$10$^{-5}$\,M$_\odot$\,yr$^{-1}$, 
a stellar radius of 4$\times$10$^{13}$\,cm, a photospheric temperature of 2330\,K, and a distance of 120\,pc (see \citealt{Agundez2009} and references therein). 
We assumed that the rotational levels of SiC$_2$ are thermalized. This approximation is adopted 
owing to the lack of information on the collision rate coefficients 
for high energy levels of SiC$_2$, and is justified by the fact that the lines observed with HIFI 
are formed in 
the warm and dense inner regions of the CSE. Furthermore, IR pumping of the lines was not included in our model.
We find that adopting an SiC$_2$ abundance of 2$\times$10$^{-7}$ relative to 
H$_2$ in the inner envelope reproduces the SiC$_2$ line profiles and intensities observed 
with HIFI reasonably well (see Fig.\,4). The largest deviation between modeled and observed line profiles is 30\% for the weakest lines.
In good agreement with the abundance derived from the rotational diagram, we adopted an initial SiC$_2$ 
abundance of 2$\times$10$^{-7}$ at a radius of 10$^{15}$\,cm, and followed the chemical composition of the gas as it expands  out 
to the outer layers of the envelope (see Fig.\,3). We find that the SiC$_2$ abundance is slightly enhanced in 
the outer envelope up to a value of 5$\times$10$^{-7}$
(Fig.\,3). Provided that the rapid reaction at low temperatures between Si and C$_2$H$_2$ yields 
SiC$_2$ \citep{Canosa2001}, this reaction is the main source of this enhancement, along with the reaction 
between Si and C$_2$H, which is assumed to occur at a rate of 10$^{-10}$ cm$^3$ s$^{-1}$. 
The uncertainties on the reaction rates of the former reaction are, however, still large.
The reaction of Si$^+$ and C$_2$H also provides a formation route to SiC$_2$ through a sequence of ion-neutral reactions. The chemical model 
indicates that SiC$_2$ is finally photodissociated to form SiC, which is not detected in our HIFI spectral survey, 
although it has been observed before in the outer envelope layers \citep{Cernicharo1989}.

The present work confirms that SiC$_2$, together with SiS and SiO, is one of the major 
silicon carriers in the inner envelope of IRC\,+10216. 
SiC$_2$ is likely formed under thermochemical equilibrium conditions near the photosphere and it may also play an important role 
in silicon carbide dust formation in the ejecta. The full HIFI line survey of IRC\,+10216 will be presented 
in a series of forthcoming papers.

\begin{figure}
\includegraphics[angle=-90,width=8.5cm]{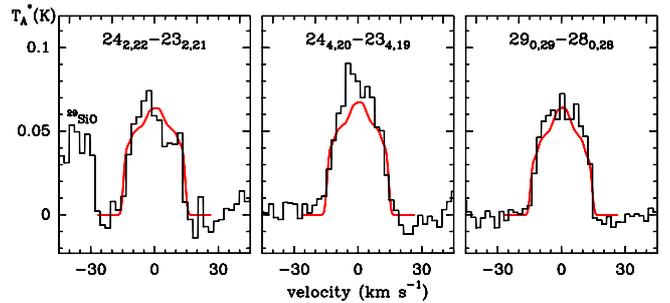}
\caption{Comparison between the line profiles of three
rotational transitions of SiC$_2$ as observed with HIFI (black
histograms) and as calculated with the radiative transfer model
(red lines) using the abundance profile shown in red in Fig.\,3.}
\label{fig-lines-model} \vspace{-0.1cm}
\end{figure}

\begin{acknowledgements}
HIFI has been designed and built by a consortium of institutes and university 
departments from across Europe, Canada, and the United States (NASA) under the leadership of 
SRON, Netherlands Institute for Space Research, Groningen, The Netherlands, and with major 
contributions from Germany, France and the US. Consortium members are Canada: CSA, U. Waterloo;
France: CESR, LAB, LERMA, IRAM; Germany: KOSMA, MPIfR, MPS; Ireland: NUI Maynooth; Italy: ASI, 
IFSI-INAF, Osservatorio Astrofisico di Arcetri-INAF; Netherlands: SRON, TUD; Poland: CAMK, CBK; Spain: 
Observatorio Astron\'omico Nacional (IGN), Centro de Astrobiologia (INTA-CSIC); Sweden: 
Chalmers University of Technology 
- MC2, RSS \& GARD, Onsala Space Observatory, Swedish National Space Board,  Stockholm University - 
Stockholm Observatory; Switzerland: ETH Zurich, FHNW; USA: CalTech, JPL, NHSC. MG and EF acknowledge 
the support from the Centre National de Recherche Spatiale (CNES). JC, MA, JRG, JRP, CSC, 
and FD thank the Spanish MICINN for funding support under grants AYA2006-14876, AYA2009-07304, and 
CSD2009-00038. LD 
and EDB acknowledge financial support from the Fund for Scientific Research - Flanders (FWO). H.S.P.M. is
grateful for support by the Bundesministerium f\"ur Bildung und Forschung (BMBF) administered through 
the Deutsches
Zentrum f\"ur Luft- und Raumfahrt (DLR), whose support was aimed in particular at maintaining the CDMS.
\end{acknowledgements}

\Online
\begin{appendix}
\section{SiC$_2$ Spectroscopy}
\begin{table}\caption{Observed SiC$_2$ transitions J$^u$$_{Ka,Kc}$-J$^l$$_{Ka,Kc}$. }\label{SiC2-lines}
\begin{tabular}{c|c|r|c|c|c}
\hline\hline\\[-2ex]
Transition    & Obs. Freq. & o-c &$S$& $E_{\rm{l}}$/k&    $I_{\varv}$\\
J$^u$$_{Ka,Kc}$-J$^l$$_{Ka,Kc}$  & (MHz)    & (MHz)&    & (K)& (K km/s)\\
\hline\\[-2ex]
24$_{\,\,2,22}$-23$_{\,\,2,21}$ &557109.0(1.5)&    1.43& 23.5 & 328.2 &  1.42( 3)\\
24$_{12,13}$-23$_{12,12}$       &557600.0(1.5)&    0.27& 18.0 & 589.7 &  0.32( 3)\\
24$_{12,12}$-23$_{12,11}$       &557600.0(1.5)&    0.27& 18.0 & 589.7 &  0.32( 3)\\
24$_{10,15}$-23$_{10,14}$       &562069.0(5.0)& $-$1.09& 19.9 & 505.2 &  0.45(11)\\
24$_{10,14}$-23$_{10,13}$       &562069.0(5.0)& $-$1.09& 19.9 & 505.2 &  0.45(11)\\
24$_{\,\,8,17}$-23$_{\,\,8,16}$ &566903.0(6.0)&    7.29& 21.4 & 436.6 &  0.74( 5)\\
24$_{\,\,8,16}$-23$_{\,\,8,15}$ &566912.0(6.0)&    9.90& 21.4 & 436.6 &  0.74( 5)\\
24$_{\,\,4,21}$-23$_{\,\,4,20}$ &568464.0(1.5)& $-$0.55& 23.3 & 346.2 &  1.25( 2)\\
25$_{16,10}$-24$_{16,\,\,9}$    &571113.0(6.0)&    2.06& 14.8 & 832.7 &  0.14( 3)\\
25$_{16,\,\,9}$-24$_{16,\,\,8}$ &571113.0(6.0)&    2.06& 14.8 & 832.7 &  0.14( 3)\\
23$_{\,\,4,19}$-22$_{\,\,4,18}$ &571141.0(1.5)&    2.43& 22.4 & 324.6 &  1.46( 2)\\
24$_{\,\,6,19}$-23$_{\,\,6,18}$ &572966.0(1.5)&    2.02& 22.5 & 384.1 &  0.90( 2)\\
27$_{\,\,0,27}$-26$_{\,\,0,26}$ &573938.0(1.5)& $-$0.07& 26.9 & 364.7 &  1.48( 2)\\
26$_{\,\,2,25}$-25$_{\,\,2,24}$ &574425.0(1.5)&    1.53& 25.7 & 361.9 &  1.27( 2)\\
24$_{\,\,6,18}$-23$_{\,\,6,17}$ &574577.0(1.5)&    3.70& 22.5 & 384.2 &  1.02( 2)\\
25$_{14,12}$-24$_{14,11}$       &576122.0(6.0)& $-$3.08& 17.2 & 716.6 &  0.13( 3)\\
25$_{14,11}$-24$_{14,10}$       &576122.0(6.0)& $-$3.08& 17.2 & 716.6 &  0.13( 3)\\
25$_{\,\,2,23}$-24$_{\,\,2,22}$ &577234.0(1.5)&    0.69& 24.5 & 355.0 &  1.39( 2)\\
25$_{12,14}$-24$_{12,13}$       &580951.0(1.5)&    0.97& 19.3 & 616.4 &  0.31( 2)\\
25$_{12,13}$-24$_{12,12}$       &580951.0(1.5)&    0.97& 19.3 & 616.4 &  0.31( 2)\\
25$_{10,16}$-24$_{10,15}$       &585735.0(1.5)& $-$0.44& 21.0 & 532.2 &  0.44( 2)\\
25$_{10,15}$-24$_{10,14}$       &585735.0(1.5)& $-$0.44& 21.0 & 532.2 &  0.44( 2)\\
25$_{\,\,8,18}$-24$_{\,\,8,17}$ &590972.0(2.5)&    0.41& 22.5 & 463.8 &  0.66( 2)\\
25$_{\,\,4,22}$-24$_{\,\,4,21}$ &590814.0(1.5)& $-$1.58& 24.3 & 373.5 &  1.09( 1)\\
25$_{\,\,8,17}$-24$_{\,\,8,16}$ &590990.0(2.5)&    3.89& 22.5 & 463.8 &  0.66( 2)\\
28$_{\,\,0,28}$-27$_{\,\,0,27}$ &594745.0(1.5)& $-$0.57& 27.9 & 392.2 &  1.46( 3)\\
27$_{\,\,2,26}$-26$_{\,\,2,25}$ &595235.0(1.5)&    2.63& 26.7 & 389.4 &  1.43( 3)\\
24$_{\,\,4,20}$-23$_{\,\,4,19}$ &596691.0(1.5)&    0.51& 23.4 & 352.0 &  1.76( 2)\\
25$_{\,\,6,20}$-24$_{\,\,6,19}$ &597462.0(5.0)& $-$0.07& 23.6 & 411.6 &  0.86( 4)\\
26$_{\,\,2,24}$-25$_{\,\,2,23}$ &597477.0(2.5)& $-$0.93& 25.5 & 382.7 &  1.10( 4)\\
26$_{14,13}$-25$_{14,12}$       &599184.0(4.0)& $-$9.03& 18.5 & 744.2 &  0.14( 2)\\
26$_{14,12}$-25$_{14,11}$       &599184.0(4.0)& $-$9.03& 18.5 & 744.2 &  0.14( 2)\\
25$_{\,\,6,19}$-24$_{\,\,6,18}$ &599914.0(1.5)&    0.11& 23.6 & 411.8 &  0.85( 2)\\
26$_{12,14}$-25$_{12,13}$       &604324.0(6.0)&    7.38& 20.5 & 644.3 &  0.38( 3)\\
26$_{12,15}$-25$_{12,14}$       &604324.0(6.0)&    7.38& 20.5 & 644.3 &  0.38( 3)\\
26$_{10,16}$-25$_{10,15}$       &609433.0(1.5)& $-$0.94& 22.2 & 560.3 &  0.47( 1)\\
26$_{10,17}$-25$_{10,16}$       &609433.0(1.5)& $-$0.94& 22.2 & 560.3 &  0.47( 1)\\
26$_{\,\,4,23}$-25$_{\,\,4,22}$ &612914.0(1.5)& $-$0.82& 25.3 & 401.8 &  1.09( 1)\\
26$_{\,\,8,19}$-25$_{\,\,8,18}$ &615115.0(4.0)& $-$0.21& 23.6 & 492.2 &  0.78( 2)\\
26$_{\,\,8,18}$-25$_{\,\,8,17}$ &615139.0(4.0)&    2.45& 23.6 & 492.2 &  0.70( 2)\\
27$_{16,12}$-26$_{16,11}$       &616666.0(6.0)& $-$1.46& 17.6 & 888.7 &  0.06( 1)\\
27$_{16,11}$-26$_{16,10}$       &616666.0(6.0)&    5.55& 17.6 & 888.7 &  0.06( 1)\\
29$_{\,\,0,29}$-28$_{\,\,0,28}$ &615547.0(1.5)&    5.55& 28.9 & 420.8 &  1.55( 2)\\
27$_{\,\,2,25}$-26$_{\,\,2,24}$ &617833.0(1.5)&    0.71& 26.5 & 411.3 &  1.15( 2)\\
25$_{\,\,4,21}$-24$_{\,\,4,20}$ &621588.0(1.5)& $-$1.38& 24.5 & 380.6 &  1.36( 2)\\
26$_{\,\,6,21}$-25$_{\,\,6,20}$ &621960.0(1.5)&    0.96& 24.7 & 440.3 &  0.91( 2)\\
27$_{14,14}$-26$_{14,13}$ &      622264.0(6.0)& $-$1.00& 19.8 & 773.0 &  0.11( 1)\\
27$_{14,13}$-26$_{14,12}$ &      622264.0(6.0)& $-$1.00& 19.8 & 773.0 &  0.11( 1)\\
26$_{\,\,6,20}$-25$_{\,\,6,19}$ &625591.0(1.5)& $-$1.43& 24.7 & 440.6 &  1.06( 2)\\
27$_{12,16}$-26$_{12,15}$ &      627699.0(1.5)& $-$0.23& 21.7 & 673.3 &  0.34( 1)\\
27$_{12,15}$-26$_{12,14}$ &      627699.0(1.5)& $-$0.23& 21.7 & 673.3 &  0.34( 1)\\
27$_{10,17}$-26$_{10,16}$ &      633168.0(1.5)& $-$0.26& 23.3 & 589.6 &  0.45( 2)\\
27$_{10,18}$-26$_{10,17}$ &      633168.0(1.5)& $-$0.26& 23.3 & 589.6 &  0.45( 2)\\
27$_{\,\,4,24}$-26$_{\,\,4,23}$ &634776.0(1.5)& $-$2.96& 26.3 & 431.2 &  1.15( 3)\\
30$_{\,\,0,30}$-29$_{\,\,0,29}$ &636346.0(1.5)&    0.01& 29.9 & 450.3 &  1.25( 4)\\
\hline\\[-2ex]
\end{tabular}
Notes: The first column lists the 
quantum numbers of the transitions. $o-c$ corresponds to observed minus
calculated frequencies.
$S$ is the line strength and $E_{\rm{l}}$ is the energy of the lower level 
of the transition. 
$I_{\varv}=\int{T_Ad\varv}$ is the integrated intensity.
Values in parentheses are uncertainies in units of the last digits
\end{table}

Silacyclopropynylidene, SiC$_2$, is a triangular molecule with a dipole moment of 2.393~(6)~D                       
\citep{SiC2_1-0_dip_1989} along the $a$-axis. $R$-branch transitions with $\Delta K_a = 0$ are                      
the strongest ones. The molecule is fairly asymmetric, $\kappa = (2B - A - C)/(A - C) = -0.7117$;                   
therefore, $Q$-branch transitions with $\Delta K_a = 0$ and transitions with $\Delta K_a = \pm2$                    
also have considerable intensities. Measurements of such transitions, even at low frequencies, may                  
improve predictions of $\Delta K_a = 0$ $R$-branch transitions at higher frequencies. SiC$_2$ has                   
a low-lying asymmetric bending mode $\nu_3$, this may lead to non-negligible changes in the dipole                  
moment with $K_a$. Only transitions with even $K_a$ are allowed because of the spin statistics associated           
with the two equivalent C nuclei.                                                     
The rotational spectrum of the main isotopic species of SiC$_2$ has been studied to some extent.
\citet{Thaddeus1984} reported the first detection of this species
towards IRC\,+10216 based on laboratory measurements of several rotational lines. 
\citet{Cernicharo1986}
reported the detection of $^{29}$SiC$_2$ and $^{30}$SiC$_2$ based on astronomical observations. 
\citet{SiC2_1-0_dip_1989} recorded the $J = 1 - 0$ transitions of the three silicon isotopologs of                  
SiC$_2$. The detection of Si$^{13}$CC in space was reported by \citet{Cernicharo1991} based on astronomical and
laboratory observations. In the same paper they reported tens of SiC$_2$, $^{29}$SiC$_2$,
$^{30}$SiC$_2$, and Si$^{13}$CC lines detected in IRC\,+10216 with frequency accuracies ranging
between 0.2 and 1.0 MHz.
At about the same time, \citet{SiC2_rot_1989} reported 34 transitions recorded between 90 and              
370\,GHz. Even though the number of parameters (15) was large,          
the data were only reproduced to within four times the quoted uncertainties. 
\citet{He2008} found that the data     
could be reproduced almost within the reported uncertainties if, instead 
of Watson's $A$-, the $S$-reduction was      
used with two more parameters plus an additional one that was estimated. To obtain a balanced fit, the    
uncertainties from \citet{SiC2_rot_1989} were multiplied by 1.5. The resulting fit was the basis of               
the CDMS\footnote{We made use of the CDMS \citep[][http://www.cdms.de]{CDMS01,CDMS05}                               
and JPL \citep[][http://spec.jpl.nasa.gov]{jpl-catalog98} molecular data bases in this work.} 
catalog entry \citep{CDMS01,CDMS05}. 

The issue of the uncertainties reported for the laboratory lines of Gottlieb et al. and their 
choice of parameters to fit the data has already been discussed in the He et al. (2008) paper. 
There it is pointed out that, despite a large number of spectroscopic parameters used in 
the fit, the experimental data was only reproduced within four times the reported uncertainties. 
He et al. were able to reproduce the data much better by switching from the A reduction 
(which was providing negative energies for J$>$25 using the Gottlieb et al. parameters) to 
the S reduction and by increasing the number of parameters somewhat. However, they were 
only able to reproduce the laboratory data within 1.5 times the experimental uncertainties. 
This finding suggests that the experimental error estimates may have been too optimistic, at 
least as long as no model is available to reproduce the experimental data better than the 
model of He et al. Moreover, the increase in experimental uncertainties by 50\% may well be too 
little because a rather large set of spectroscopic parameters was needed to fit a rather small 
set of experimental lines of this admittedly non rigid molecule. As discussed below, we found 
in the present investigation that the modified uncertainties by He et al. are appropiate 
for our present, more larger data set.
The strong transitions are predicted quite well up to about 500\,GHz,          
but the quality of the prediction deteriorates rapidly at higher frequencies, as it turned out, in particular for   
lower values of $K_a$. Therefore, the SiC$_2$ transition frequencies from the present HIFI observations were        
subjected to a combined fit with the laboratory data.                                                               
Table\,\ref{SiC2-spec-parameters} compares the present parameters                                                   
with those from \citet{He2008}. All parameters were                                                                 
improved, especially the ones that depend particularly on $J$. Even though the                                     
parameter $L_J$ was newly introduced, the uncertainty of $H_J$ also decreased. 
In most cases, the parameter values differ only slightly from the previous ones, validating the previous model. 
Some changes outside the uncertainties are likely caused by including $L_J$ in the fit. \citet{He2008} 
have already found that 
only slight modifications of the parameter values are required to fit isotopic data from their,
and previous astronomical 
\citep{Cernicharo1991} and laboratory data \citep{SiC2_1-0_dip_1989}.
The SiC$_2$ lines detected in the    
HIFI 1b survey, their uncertainties, and residuals are listed in Table\,\ref{SiC2-lines}.                            
                                                                                                                    
\begin{table}
\begin{center}
\caption{Spectroscopic parameters$^a$ (MHz) for SiC$_2$ obtained in the present investigation 
compared to a previous study.}
\label{SiC2-spec-parameters}
\begin{tabular}{lrr}
\hline \hline
Parameter & This work & He et al. 2008\\
\hline
$A-(B+C)/2$               &  40674.000\,(81)    &  40674.191\,(100)   \\
$(B+C)/2$                 &  11800.11932\,(84)  &  11800.11861\,(99)  \\
$(B-C)/4$                 &    678.41182\,(76)  &    678.41301\,(130) \\
$D_K$                     &   $-$1.3429\,(185)  &   $-$1.2880\,(227)  \\
$D_{JK}$                  &      1.624141\,(59) &      1.624219\,(88) \\
$D_J \times 10^3$         &   $-$1.1651\,(80)   &   $-$1.1845\,(115)  \\
$d_1 \times 10^3$         &   $-$2.4371\,(53)   &   $-$2.4392\,(111)  \\
$d_2 \times 10^3$         &   $-$7.1766\,(30)   &   $-$7.2002\,(80)   \\
$H_{KJ} \times 10^6$      &    669.68\,(104)    &    670.39\,(151)    \\
$H_{JK} \times 10^6$      & $-$138.26\,(12)     & $-$138.66\,(39)     \\
$H_J \times 10^6$         &      0.9810\,(187)  &      1.0079\,(201)  \\
$h_1 \times 10^9$         & $-$193.0\,(101)     & $-$180.2\,(247)     \\
$h_2 \times 10^9$         & $-$578.1\,(84)      & $-$513.3\,(301)     \\
$h_3 \times 10^9$         &    185.8\,(67)      &    157.0\,(93)      \\
$L_{KKJ} \times 10^9$     & $-$135.4\,(69)      & $-$138.5\,(80)      \\
$L_{JK} \times 10^9$      &    14.68\,(175)     &    14.34\,(266)     \\
$L_{JJK} \times 10^9$     &  $-$12.11\,(28)     &  $-$11.06\,(78)     \\
$L_J \times 10^{12}$      &    138.9\,(166)     &       $-$           \\
$P_{KKKJ} \times 10^{12}$ &     20.$^b$         &     20.$^b$         \\
\hline
\end{tabular}\\[2pt]
\end{center}
$^a$ Watson's $S$-reduction was used in the representation $I^r$. 
Numbers in parentheses are 1\,$\sigma$ uncertainties in units of the 
least significant figures.\\
$^b$ Kept fixed to estimated value.
\end{table}
\end{appendix}
\end{document}